\documentstyle[sprocl]{article}
\input{psfig}
\def\Journal#1#2#3#4{{#1} {\bf #2}, #3 (#4)}
\def\be{\begin{equation}}
\def\ee{\end{equation}}
\def\bea{\begin{eqnarray}}
\def\eea{\end{eqnarray}}

\def\k{\bf k}

\newcommand{\lexp}{\mathop{\langle}}
\newcommand{\rexp}{\mathop{\rangle}}
\def\apj{\em Astrophys. J.}

\begin{document}
\title{Understanding Gravitational Clustering with Non-Linear
Perturbation Theory}
\author{Rom\'an Scoccimarro}
\address{CITA, McLennan Physical Labs, 60 St George Street, 
Toronto ON M5S 3H8}

\maketitle\abstracts{I discuss new results concerning the 
evolution of the bispectrum 
due to gravitational instability from gaussian initial conditions 
using one-loop perturbation theory (PT). Particular attention is paid to the 
transition from weakly non-linear scales to
the non-linear regime at small scales. Comparison with 
numerical simulations is
made to assess the regime of validity of the perturbative approach. 
}

\section{Introduction}

This work is based on results to be reported in [1]. 
Here, I briefly present 
results on the one-loop corrections to the bispectrum 
and compare them to numerical simulations for CDM initial 
spectra.\footnote{with  
$\Omega=1$, $\Gamma= 0.25$. The 
simulation data is publically available through the Hydra Consortium 
Web page ({\sf http://coho.astro.uwo.ca/pub/consort.html}).}
In particular, we work in terms of the
hierarchical amplitude $Q$ defined from the bispectrum, 
$ B({\k}_1, {\k}_2) $, and power spectrum, $P(k)$, as follows
(superscripts denote, tree-level, one-loop PT, and so on):
\be
Q \equiv { {B({\k}_1,{\k}_2)} \over \Sigma({\k}_1,{\k}_2)} =
\frac{B^{(0)}{({\k}_1,{\k}_2)}+B^{(1)}{({\k}_1,{\k}_2)}+\ldots}{
\Sigma^{(0)}{({\k}_1,{\k}_2)}+\Sigma^{(1)}{({\k}_1,{\k}_2)}+\ldots} =
Q^{(0)}+Q^{(1)}+\ldots  ,
\label{q}
\ee
with: 
\be
\lexp {\delta({\k}) \delta({\k}') \rexp = \delta_D({\k}+{\k}')}\   P(k) ,
\ee
\be
{\lexp \delta({\k}_1) \delta({\k}_2) \delta({\k}_3) \rexp
= \delta_D({\k}_1+{\k}_2+{\k}_3) }\  B({\k}_1,{\k}_2) .
\ee
\be
{\Sigma({\k}_1,{\k}_2)} \equiv P(k_1) \ P(k_2) + P(k_1) \ P(k_3) + P(k_2) \ P(k_3)
\ee
where we have used that the bispectrum is 
defined for closed triangle configurations,
$ \sum_{i=1}^3 {\k}_i = 0 $. The perturbative quantities in Eq.~(\ref{q})
can be calculated from the standard machinery of PT.\cite{SCFFHM97,Sco97} 
In the following, we consider $Q$ for configurations where $k_1/k_2=2$,
as a function of $\theta$, the angle between $\hat{{\k}}_1$ and $\hat{{\k}}_2$.

%\clearpage

\begin{figure}[p] 
\psfig{figure=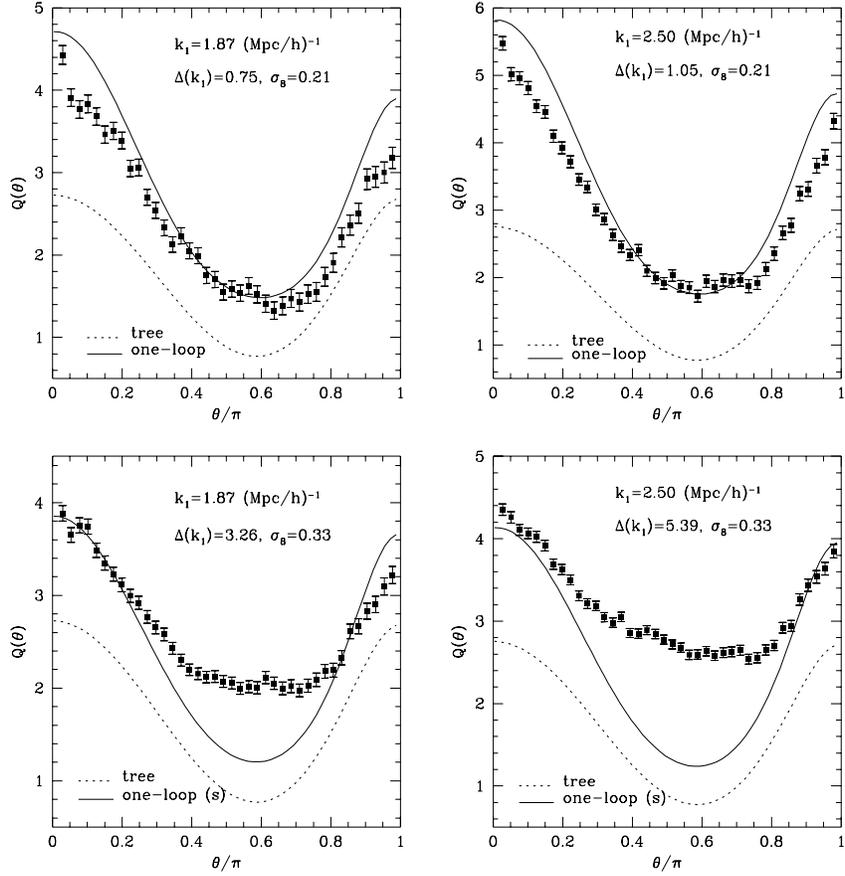,height=13. truecm,width=12.5 truecm} 
\caption{The 
hierarchical amplitude $Q$ for triangle configurations with 
$k_{1}/k_{2}=2$
as  a function of the angle $\theta$ between  $\hat{{\k}}_{1}$ and 
$\hat{{\k}}_{2}$ in the Hydra CDM numerical simulations (symbols), 
tree-level PT (dotted lines) and one-loop PT (solid lines).}
\label{fig1}
\end{figure}

\section{Results}

Figure~\ref{fig1} shows $Q$ in numerical simulations compared to 
tree-level PT and one-loop PT.  Error bars in these plots 
are estimated from the number of independent Fourier modes
contributing to each configuration 
assuming gaussianity.\cite{SCFFHM97} The degree of 
non-linearity in each case
can be inferred from the dimensionless power spectrum,
$\Delta(k) \equiv 4\pi k^3 P(k)$. 
In the top panels ($\sigma_8=0.21$), 
we note a clear deviation of the N-body results 
from the tree-level PT prediction for $Q$, and a good agreement with
the one-loop correction. At this stage of non-linear evolution, 
the dynamics is dominated by large-scale power and therefore an 
enhancement of Q at collinear configurations ($\theta=0,\pi$) 
develops.\cite{Sco97}
In the bottom panels ($\sigma_8=0.33$), where already 
$\Delta (k_1) >1$, we use the ratio of one-loop quantities in 
Eq.~(\ref{q}) (denoted as ``one-loop (s)'' in Fig.~\ref{fig1}) 
for the one-loop prediction.\cite{Sco97}
We see very good agreement for configurations close to
collinear, and a progressively flattening of $Q(\theta)$
as we look at smaller scales. The flattening is due to configurations
close to equilateral becoming more probable due to random motions at
small scales.\cite{SCFFHM97} At even more non-linear scales, $Q$ becomes
configuration independent, in rough agreement with the hierarchical
ansatz for the three-point function.\cite{SCFFHM97}

\section*{Acknowledgments} 

This work is based on a project in collaboration with 
S.~Colombi, J.N.~Fry, J.A.~Frieman, E.~Hivon, \& A.~Mellot.
I would like in addition to thank F.~Bernardeau, 
E. Gazta\~{n}aga, B.~Jain, R.~Juszkiewicz, C.~Murali for 
conversations, and especially Hugh Couchman  
for numerous helpful discussions.
The CDM simulations analyzed in this work were obtained
from the data bank of cosmological N-body simulations provided by the
Hydra consortium ({\sf http://coho.astro.uwo.ca/pub/data.html}) and produced
using the Hydra N-body code.\cite{CTP95}


\section*{References} 
\begin{thebibliography}{99} 

\bibitem{SCFFHM97}
R. Scoccimarro, S. Colombi, J.N. Fry, J.A. Frieman, E. Hivon, and
A.~Melott,  in preparation (1997). 

\bibitem{Sco97}
R. Scoccimarro,  submitted to {\apj}, astro-ph/9612207, (1996).

\bibitem{CTP95}
H. Couchmann, P.A. Thomas, and  F. Pearce, \Journal{\apj}{452}{797}{1995}. 






\end{thebibliography}
\end{document}